\documentclass[12pt]{article}
\usepackage{amsmath}
\usepackage{amsfonts}
\usepackage{graphicx}
\usepackage{xcolor}
\graphicspath{ {../images/} }
\usepackage[margin=1in,top=1.5in,bottom=1.5in]{geometry}

\begin{document}

\title{Critical Dynamics of Random Surfaces:\\
Time Evolution of Area and Genus}

\author{\\[0pt] Christof Schmidhuber\footnote{christof@schmidhuber.ch}\\ [15pt]
Zurich University of Applied Sciences, Switzerland\\ [5pt]
\\  [0pt]
}

\maketitle

\begin{abstract}\vspace{3mm}
Conformal field theories with central charge $c\le1$ on random surfaces have been extensively studied in the past.
Here, this discussion is extended from their equilibrium distribution to their critical dynamics. 
This is motivated by the conjecture that these models describe the time evolution of certain social networks that are self-driven to a critical point.
This paper focuses on the dynamics of the overall area and the genus of the surface.
The time evolution of the area is shown to follow a Cox Ingersol Ross process. 
Planar surfaces shrink, while higher genus surfaces grow to a size of order of the inverse cosmological constant.
The time evolution of the genus is argued to lead to two different phases, dominated by (i) planar surfaces, 
and (ii) ``foamy'' surfaces, whose genus diverges.
In phase (i), which exhibits critical phenomena, time variations of the order parameter are approximately t-distributed with 4 or more degrees of freedom. 
\end{abstract}

\newpage

\section{Introduction}

Conformal field theories with central charge $c\le1$ on random surfaces have been extensively studied in string theory.
Their continuum field theory was developed in \cite{poly1,kpz,DDK,DDK2}, 
while the dual matrix model approach \cite{brezin,brezin2,matrix,matrix2,kaza,kaza2} gave insights into the sum over surface topologies \cite{mxm,mxm2,mxm3,migdal,kleb}.
These models can be viewed as noncritical string theories in a $c+1$-dimensional target space, where the 
random surface represents the string world-sheet, and the world-sheet conformal factor acts as a new embedding dimension. 
For $c>1$, the random surfaces are unstable and are believed to degenerate to branched polymers.\\

As far as the author knows, this discussion has been restricted to the static limit (although there has beeen a stochastic quantization approach \cite{yon}). 
This is analogous to modeling the equilibrium distribution of water and steam at its critical point, independently of time.
However, if one wants to compute dynamic properties such as the correlation between the steam pressure at different times,
one must go beyond this static limit and study the critical dynamics \cite{hohenberg} of water and steam. Analogously,
in order to study the time evolution of random surfaces, we here extend their theory include its critical dynamics.\\
  
Within string theory, there is no obvious reason to study the critical dynamics of random surfaces, as world-sheet time is already one of the two dimensions of the surfaces.
There is no need to extend the world-sheet by a third, nonrelativistic time dimension.
However, there may be other statistical-mechanical applications of random surfaces or their dual graphs (which can be regarded as networks), 
where it is important to study their time evolution. 
This includes the dynamics of social networks, which come in a huge variety of network topologies such as trees, small-world networks,
scale-free networks, etc. \cite{watts,watts2,cimi}.\\

\begin{figure}[t!]\centering
	\includegraphics[height=5.5cm]{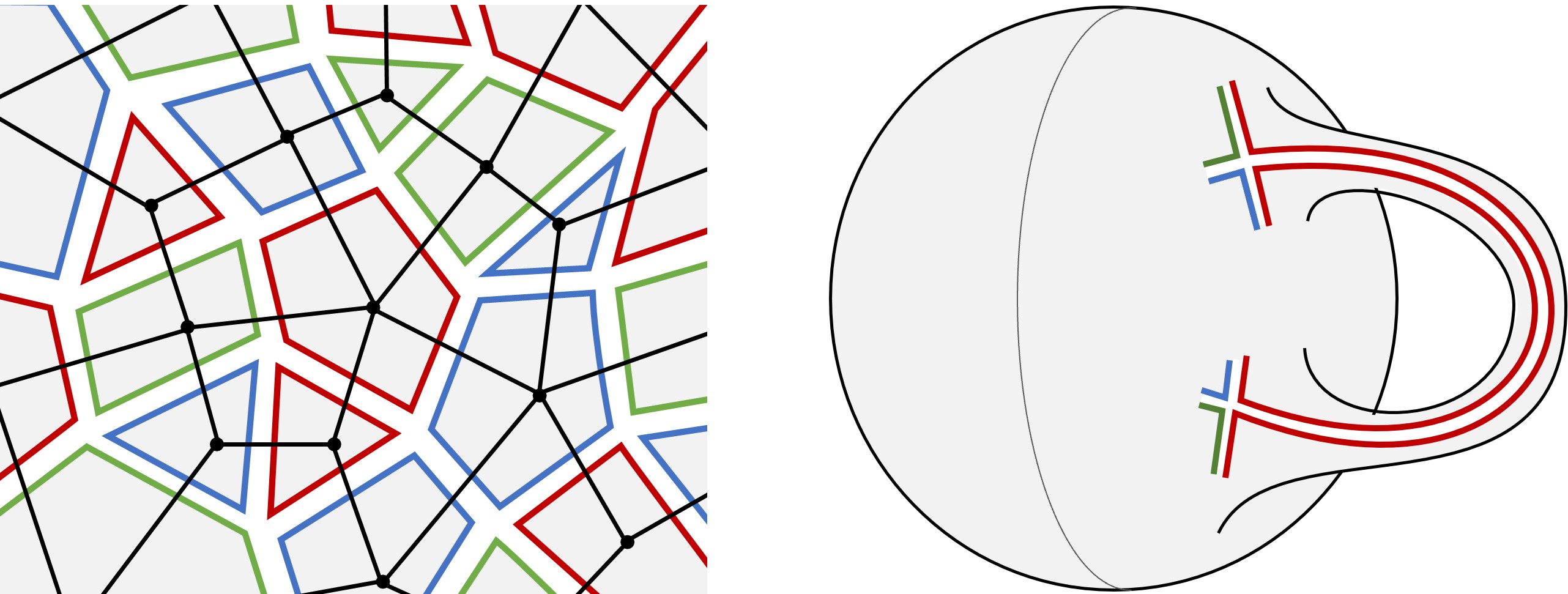}
	\caption{Left: Planar diagram of a matrix model, viewed as a network with colored double lines as links. Its dual graph is a curved surface. 
         Right: long range links increase its genus.}
\end{figure}

Most of these networks are highly connected, so one expects the large-scale limit of any statistical mechanical model that lives on the networks
 to be described by mean field theory.
By contrast, in this note we focus on ensembles of random networks that have a non-trivial continuum limit, 
i.e., that are described at large scales by some $D$-dimensional 
renormalizable field theory, which can explain observed analogies with critical phenomena.\\

We expect the fields of such a continuum theory to include gravity, which represents local fluctuations of the network's connectivity, i.e., of the geometry and topology of the dual graphs.
This is illustrated in fig. 1 at the example of the Feynman diagrams of the well-known "matrix models" \cite{brezin,brezin2} of real symmetric $N\times N$ matrices.
We interpret them as network graphs, whose links are double lines. Four links meet at each vertex, and each line has $N$ possible colors. 
The dual graph (solid black lines) is a discretized surface made up of quadrangles, 
which can be deformed into squares by leaving the two-dimensional embedding space. 
At a corner where $k$ squares meet, the curvature is proportional to $4-k$.
The ensemble of networks is thus dual to an ensemble of discretized curved surfaces. 
Long-range links, which are suppressed in the matrix model partition function by a factor of $1/N^2$ for each link, correspond to topology fluctuations.
We refer to \cite{brezin} for an introduction.\\
 
The requirement that any nontrivial continuum limit of random networks, if it exists, must include a renormalizable theory of gravity
restricts us to dimensions $D=0, 1$ or $2$ (however, see, \cite{ambjorn} for work on $D>2$).
The trivial case $D=0$ corresponds to mean field theory.
An example of the more interesting case $D=1$ are branched polymers. In this work, we focus on the most interesting case $D=2$, i.e., random surfaces,
which also includes cases where the surfaces are believed to degenerate to branched polymers. \\

This work is organized as follows. Section 2 reviews the relevant background on random surfaces in conformal gauge and on critical dynamics, then combines both. 
In section 3, we study the critical dynamics in "minisuperspace approximation", which means that we treat only the overall surface area as a dynamic variable.
We find that its time evolution follows a Cox-Ingersol-Ross process \cite{cir,cir2}. 
Planar surfaces shrink linearly in physical time, while higher genus surfaces grow until their growth is halted by the cosmological constant.
We also discuss the analogous time evolution of surfaces with operator insertions.\\

In section 4, we also allow the genus of the random surface to be dynamical.
We conclude that there are two different regimes into which the ensemble of random surfaces can evolve in time. They are dominated by
(i) small planar surfaces, and (ii) ''foamy'' surfaces whose genus diverges, corresponding to a condensation of handles.
Our conclusions about regime (ii) are based on nonperturbative results from the matrix models \cite{mxm,mxm2,mxm3,migdal}. \\

In section 5, we study time variations of the order parameter, which we call ``returns''. 
We find that their distribution is not Gaussian, but a generalized hyperbolic distribution in minisuperspace approximation. 
In regime (i), it has power-law tails with a tail index of 4 or more. 
The volatility of returns is not constant in time, but displays clusters and spikes. 
A follow-up paper \cite{companion} extends this analysis beyond the minisuperspace approximation.
\\

Section 6 summarizes the results and comments on the original motivation for this work, namely
empirical observations of analogies between financial markets and critical phenomena \cite{mant,mant2,bouch1,schmi1}.
It has recently been proposed to explain them by a lattice gas model of the markets \cite{schmi2}, where the
lattice represents the social network of investors.
While the current paper is independent of this potential application, it provides a basis for working out its predictions for random networks in the future.\\

\section{Field Theory Formulation}

In this section, we first briefly summarize aspects of the theory of random surfaces and of critical dynamics that are relevant for this paper, and then combine them.

\subsection{Brief Review of Random Surfaces}

We consider a two-dimensional Euclidean field theory on a random surface with coordinates $\sigma\equiv(\sigma_1,\sigma_2)$. 
This could, e.g., be a scalar field theory with field $x(\sigma)$. On a fixed surface with metric $g_{\alpha\beta}$, its classical action is
\begin{equation}
S_{CFT}[g,x]= \int d^2\sigma\ \sqrt{\det g}\ \{{1\over2}g^{ij}\partial_i x\partial_j x+V(x) \},\label{matter}
\end{equation}
where the potential is, e.g., of the form $V(x)=r/2\cdot x^2+g/24\cdot x^4$ for the Ising model. 
For $r=0$, the theory flows to a renormalization group fixed point (a "conformal field theory") with some critical coupling $g=g^*$, where it describes the critical point of the Ising model,
and the field $x$ has anomalous dimension $\eta/2=1/8$.  
More generally, we will consider as conformal field theories the "unitary minimal models" \cite{BPZ} with central charges
$$c=1-{6\over m(m+1)},\ \ m\in\{3,4,5,...\}.$$
In the Landau-Ginzburg description, they correspond to potentials of the form $x^{2m-2}$. 
Their operators of definite scaling dimension (primary fields) are labelled by two integers $p\ge q\ge1$ (with $p\le m-1, q\le m$). 
In particular, the operator $\Phi$ with $p=q=2$, which we use as an order parameter, has anomalous dimension
$$\text{dim}(\Phi)={\eta\over2}\equiv\Delta\equiv2h_{22}={3\over2m(m+1)}.$$
The case $m=3$ corresponds to the Ising model, with $\Phi$ corresponding to the magnetization. $m=4$ corresponds to the tricritical Ising model,  $m=5$ (or rather, a subset of it) to the 3-states Potts model, and so on.
We call these models the "matter". When putting matter on a random surface, we are restricted to central charges $c\le1$, 
because for $c>1$ the surfaces turn out to become unstable (the "tachyon problem" of bosonic string theory); they are believed to degenerate to branched polymers.
 
By "putting the matter on a random surface", we mean that the two-dimensional metric becomes dynamical, i.e., the path integral includes a sum over two-dimensional metrics and topologies. 
Up to reparametrization, two-dimensional metrics can locally be written in "conformal gauge" as
\begin{equation}
g_{ij}(\sigma)=\delta_{ij}\cdot e^{\phi(\sigma)}\ \ \circ\ \ \text{Diffeomorphism},\label{conformalgauge}
\end{equation}
where $\phi$ is the "conformal factor". The topologies of closed two-dimensional surfaces are labelled by their genus $g$, the number of handles, which is related to the Euler characteristic
$$\chi=2-2g={1\over4\pi}\int d^2\sigma\sqrt{g}\ R(\sigma),$$
where $R$ is the two-dimensional curvature tensor. For a surface of genus $g>0$, the metric can only locally be reduced to the form (\ref{conformalgauge}). 
Globally, there remains an $M_g$ dimensional space of moduli $m_i$ that we must also integrate over ($M_1=2,M_{g>1}=6g-6$). Altogether, 
the partition function of a conformal field theory on random surfaces of all geni $g$ is \cite{DDK,DDK2}
\begin{eqnarray}
Z&=&\sum_{g=0}^\infty\ \prod_{i=1}^{M_g}dm_i\int D\phi\ Dx\ \exp\{-S_\text{CFT}[x]-S_G[\phi]-S_A[\phi]+\lambda\Phi[x] e^{\alpha_{22}\phi}\}\label{partition}\\
S_G&=&\int d^2\sigma\sqrt{\det \hat g}\ \{\gamma\hat R+\mu e^{\alpha\phi} \}\label{gravity}\\
S_A&=&{1\over8\pi}\int d^2\sigma\sqrt{\det \hat g}\ \hat g^{ij}\{\partial_i \phi\partial_j\phi +Q\hat R_{ij}\phi\}\label{anomaly}
\end{eqnarray}
where $\hat g$ is some auxiliary background metric, and $S_G$ is the gravitational action consisting of the Hilbert-Einstein term and a cosmological constant $\mu$. 
Being at a renormalization group fixed point, the matter theory with $\lambda=0$ is conformally invariant at the classical level, i.e., independent of $\phi$.
At the quantum level, however, the effective action $S_A$ (\ref{anomaly}) is induced by the conformal anomaly $c$ \cite{poly1}
($c$ has been absorbed in a field rescaling) with renormalization parameters $\alpha$ and $Q$.
Surfaces of genus $g$ are weighted by 
\begin{equation}
(\kappa^2e^{Q\phi_0})^{g-1}\ \ \ \text{with topological coupling constant}\ \ \kappa^2=\exp\{8\pi\gamma\}\label{topo}
\end{equation}
in the partition function, where $\phi_0$ is the spatially constant mode of $\phi$.
Model (\ref{partition}) is perturbed away from the fixed point by a small coupling constant $\lambda$,
where $\Phi[x]$ is the order parameter and $e^{\alpha_{22}\phi}$ is its so-called "gravitational dressing".

The theory must be invariant under rescalings of the arbitrarily chosen background metric $\hat g_{\alpha\beta}$. 
In particular, its conformal anomalies must cancel, and the $\lambda$-perturbation must be exactly marginal (and, i.p., have dimension $0$). 
This determines $Q, \alpha$, and $\alpha_{22}$:
\begin{equation}
3Q^2=25-c\ ,\ \ \alpha(Q-\alpha)=2\ , \ \ \alpha_{22}(Q-\alpha_{22})+\Delta=2\ ,\label{cosmo}\\
\end{equation}
as computed from conformal field theory. For the $m$-th minimal model, we get
\begin{equation}
\alpha^2={2m\over m+1}\ ,\ \ \ {Q\over\alpha}=2+{1\over m}\ ,\ \ \ 2{\alpha_{22}\over\alpha}=2-{1\over m}.\label{ddk}
\end{equation}
If all operators have dimension 2, how can there be nontrivial scaling dimensions?
Since $A=\int d^2\sigma \ e^{\alpha\phi}$ is the area (which has dimension $-2$), {\it physical} (as opposed to background) rescalings by a factor $e^{-\tau}$ 
correspond to constant shifts of the field $\phi$: 
\begin{equation}
\phi\rightarrow\phi-{2\over\alpha}\tau.\label{shift}
\end{equation}
Thus, while nothing depends on the background scale of the metric $\hat g_{\alpha\beta}$,
the physical scale dependence is encoded in the $\phi$-dependence of $\lambda_i e^{\alpha_i\phi}$. It can be expanded to higher orders in $\lambda$ \cite{schmi3,schmi4}. 
At lowest order, the ``gravitationally dressed dimension'' of $\Phi$ is thus
$2-2{\alpha_{22}/\alpha}=1/m$ (before integrating over the two-dimensional surface).\\

By shifting $\phi$, we can also infer the partition function $Z_{g,A}$ for fixed area $A$ and genus $g$:
\begin{equation}
Z_{g,A}=\langle \delta(\int  d^2\sigma\ \sqrt{\hat g}\ e^{\alpha\phi}-A)\rangle\sim A^{-1+(g-1){Q\over\alpha}}\ e^{-\mu A}.\label{fixedA}
\end{equation}
The distribution of genus-zero surfaces is not normalizable  \cite{seiberg} and dominated by surfaces with area $A\approx l^2$, where $l$ is a short-distance cutoff.
A natural way to introduce such a cutoff is to add a boundary of fixed length $l$ to the surface. Then the distribution becomes \cite{stau}
\begin{equation}
Z_{A,l}\sim e^{-l^2/A}\ A^{-{Q\over\alpha}}\ l^{-3+{Q\over\alpha}}\ e^{-\mu A}.\label{cutoff}
\end{equation}
We see from (\ref{fixedA}) that genus-1 surfaces are distributed across all sizes. Higher genus surfaces are dominated by large areas, which are eventually cut off by the cosmological constant $\mu$. Each handle comes with a factor $A^{Q/\alpha}=A^{2+1/m}$
(naively, one would have expected $A^2$, as both ends of the handle can lie anywhere on the surface). 
Thus, large surfaces are crowded with handles, while small surfaces are predominantly planar (i.e., have genus zero).

\subsection{Brief Review of Critical Dynamics}

For an in-depth review of critical dynamics, see \cite{tauber,ZJ}. Here we only mention a few aspects. 
A random walk $q(t)$ in a potential $V(q)$ is described by the stochastic differential equation
$$ \dot q(t) = -{\Omega\over2}V'(q) + \nu(t)\ \ \ \text{with "noise"}\ \langle \nu(t)\nu(t')\rangle=\Omega\delta(t-t'),$$
where the $\nu(t)$ are independent, normally distributed random variables with mean zero and variance $\Omega$.
The stochastic process dissipates to the equilibrium probability distribution
\begin{equation}
P(q)=\exp\{-V(q)\}.\label{statpot}
\end{equation}
Thus, the static limit of the stochastic process corresponds to that of a zero-dimensional quantum particle (without time) in the potential $V(q)$.
In the case $V=aq$, $q(t)$ is a Brownian motion with drift $a$, and there is no equilibrium distribution. \\

The random walk can also be treated in path integral formulation with partition function
\begin{eqnarray}
Z&=&\int D\nu(t)\ \exp\{-{1\over2\Omega}\int dt\ \nu^2\}\notag\\
&=&\int D q(t)\ \exp\{-{1\over2\Omega}\int dt\ \big[\big(\dot q+{\Omega\over2}V'(q)\big)^2-{\Omega^2\over2} V''(q)\big]\},\label{dynaction}
\end{eqnarray}  
where the last term represents the Jacobian $\det(\partial_t+{\Omega\over2}V'')$ that comes with the change of variables from $\nu(t)$ to $q(t)$.
$Z$ can be elegantly rewritten in term of auxiliary variables $\lambda(t)$ and fermionic variables $C(t),\bar C(t)$ with anti-commutator $\{\bar C(t),C(t')\}=\delta(t-t')$ and action
\begin{eqnarray}
\Rightarrow\ \ \ S&=&\int dt\ \{-{\Omega\over2}\lambda^2 +\lambda\big(\dot q+V'(q)\big)-\Omega\ \bar C\big(\partial_t+V''(q)\big)C\}\label{action}
\end{eqnarray}
It is invariant under a global supersymmetry generated by the fermionic variable $\epsilon$:
\begin{equation}
\delta q=\bar C\epsilon,\ \ \delta C=(\lambda-\dot q)\epsilon,\ \ \delta \bar C=0,\ \ \delta\lambda=\dot{\bar C}\epsilon.\label{susy}
\end{equation}

This discussion straightforwardly generalizes from a particle $q(t)$ to a $D$-dimensional field $x(\vec \sigma,t)$ in a potential $V(x)$ 
with noise $\nu(\vec \sigma,t)$, with dynamics given by the Langevin equation
\begin{equation}
\partial_tx=-{\Omega\over2}\cdot {\delta S[x]\over\delta x}+\nu\ ,\ \text{where}\ \ {\delta S[x]\over\delta x}=-\Delta x+V'(x).\label{modelA}
\end{equation}
This dynamics is called "model A". The probability distribution in the static limit is now
\begin{equation}
P[x(\vec \sigma)]=\exp\{-\int d^D\vec \sigma\big[{1\over2}\partial_ix\partial^ix+V(x)\big]\}.\notag
\end{equation}
Thus, in the static limit (at large times) the system reduces to ordinary Euclidean quantum field theory in $D$ dimensions. 
We will focus on model A, as - in the absence of conserved quantities - it is known to lie in the unique dynamic universality class with this static limit.\\

The $D$-dimensional version of the supersymmetry (\ref{susy}), which acts only in time and not in space,
ensures that the structure of the action (\ref{action}) is preserved under renormalization, 
although $\Omega$ acquires an anomalous dimension. At two-loop level, it is related to the dimension $\eta$ of the field $\phi$ by
$$\text{dim}(\Omega)=(c+1)\eta \ \ \text{with}\ \ \ c+1=6\ln{4\over3}\approx1.726. $$
$\Omega$ can be absorbed in the time $t$ in the diffusion equation (\ref{modelA}). Dimension counting implies that the classical scale invariance
 $\vec \sigma\rightarrow \lambda\vec \sigma, t\rightarrow \lambda^2 t$ is modified at the quantum level to
\begin{equation}
\vec \sigma\rightarrow \lambda\vec \sigma\ \  , \ \ t\rightarrow \lambda^z t\ \ \ \text{with}\ z=2+c\cdot \eta.\label{scale}
\end{equation}
In particular, the "correlation time" $\tau$ is related to the correlation length $\xi$ by $\tau=\xi^z$, and 
two-dimensional Peierls droplets of area $A$ decay over a physical time span of order $T\sim A^{z/2}$, instead of the classical decay time $T\sim A$. 
In the case of the Ising model, $z$ has been computed up to 5 loops with the result $z\approx13/6$ \cite{night}, corresponding to $c\approx2/3$.

\subsection{Critical Dynamics of Random Surfaces}

We now apply the critical dynamics (\ref{modelA}) of model A to the minimal models on a random surface.
To this end, we combine the actions (\ref{matter},\ref{gravity},\ref{anomaly}) and introduce a time dimension $\hat t$. We call $\hat t$ the ``background time'',
as it trivially extends the two-dimensional background metric $\hat g_{ij}$ to a three-dimensional one: $\hat g_{\hat t\hat t}=1, \hat g_{\hat t i}=0$.
There are now two dynamic critical coefficients: $z$ for the matter, and $z_\phi$ for the gravitational sector. Since $\phi$ has dimension 0, $z_\phi=2$ from (\ref{scale}). 
The dynamic action for $\phi$ is derived from (\ref{gravity}, \ref{anomaly}, \ref{dynaction}):
\begin{eqnarray}
S[\phi]&=&{1\over2\Omega}\int d\hat t\int d^2x \sqrt{\vert\hat g\vert} \big\{\big[\partial_{\hat t}\phi
-{\Omega\over8\pi}\hat \Delta\phi+{\Omega\over16\pi}{Q}\hat R+{\mu\alpha\Omega\over2}e^{\alpha\phi}\big]^2
-{\mu\alpha^2 \Omega^2\over2}e^{\alpha\phi}{\color{red}}\big\}\ \ \ \label{triangle}
\end{eqnarray}
How is background time $\hat t$ related to physical time $t$ in this nonrelativistic theory?
In accordance with (\ref{scale}) and for $z=2$, {\it physical} spatial and time distances are given by
$$\vert\delta x\vert^2=e^{\alpha\phi}\vert\delta \hat x\vert^2\ ,\ \ \ \vert\delta t\vert^2=\Omega^2e^{2\alpha\phi}\vert\delta \hat t\vert^2.$$
Here, we regard $g_{tt}\equiv\Omega^2 e^{2\alpha\phi}$ as a metric component that extends the two-dimensional {\it physical} metric $g_{ij}$
to three-dimensions. We conclude that $\hat t$ and $t$ are related by 
\begin{equation}
{\partial\over \partial \hat t} t(\sigma,\hat t)= \Omega e^{\alpha\phi}\ \ \ ,\ \ \ {\partial\over \partial t}\hat t(\sigma,t)= \Omega^{-1} e^{-\alpha\phi},\ \ \ 
\label{time}
\end{equation}
Using (\ref{time}), we can also write the action in physical time with physical metric $g_{ij}=\hat g_{ij}e^{\alpha\phi}$:
\begin{eqnarray}
 S_\phi&=&{1\over2}\int dt\int d^2x\ \sqrt{\vert g\vert}\ 
\Big\{\big[\partial_t\phi+ {1\over16\pi}  \tilde R +{\mu\alpha\over2}\big]^2-{\mu\alpha^2 \over2}e^{-\alpha\phi}\Big\}. \label{phys}
\end{eqnarray}
Here, $\tilde R=e^{-\alpha\phi}({Q}\hat R-2\hat\Delta\phi)$ is the rescaled physical curvature. Note that $\Omega$ drops out of (\ref{phys}).\\

In the following, we will set $\Omega=1$ and choose a background metric $\hat g_{ij}$ with constant curvature for each genus $g$. 
We split the conformal factor $\phi(\sigma,t)=\phi_0(t)+\tilde\phi(\sigma,t)$ into the spatially constant mode (or ``zero-mode'') $\phi_0$ and the remainder $\tilde\phi$:
$$\phi_0(t)=\int_\Sigma d^2\sigma\ \phi(\sigma,t)\ \ \ ,\ \ \ \tilde\phi(\sigma,t)=\phi(\sigma,t)-\phi_0(t)\ \ \Rightarrow\ \ \int_\Sigma d^2\sigma\ \tilde\phi(\sigma,t)=0$$
 Only the zero mode ``sees'' the background charge.
The respective actions decouple for $\mu=0$:
\begin{eqnarray}
S[\phi_0]&=&{1\over2}\int d\hat t\ \big\{\dot\phi_0+{Q\over2}(1-g)\big\}^2\label{zero}\\ 
S[\tilde\phi]&=&{1\over2}\int d\hat t\ \big\{\dot{\tilde\phi}-{\Omega\over8\pi}\Delta\tilde\phi)^2\big\}^2.\notag
\end{eqnarray}
Although setting $\mu=0$ makes no sense in the static limit, as there is then no equilibrium distribution, 
it can be a useful approximation for the dynamic model far from equilibrium.

\section{Dynamics of the Area}

In this section, we discuss the dynamics of the zero mode $\phi_0$. 
That is, we work in the ``minisuperspace approximation'', where only the overall area 
$A(\hat t)\sim e^{\alpha\phi_0(\hat t)}$ is dynamical. 
In the static limit, this approximation has surprisingly yielded exact results such as the scaling behaviour (\ref{fixedA},\ref{cutoff}).
Based on this, we conjecture that the results in this section are also exact. 

\subsection{Minisuperspace Approximation}

We begin with genus zero.
Regarding $\phi_0$ as a stochastic process, its differential equation in background time $\hat t$ is easily solved for zero cosmological constant $\mu$. From (\ref{zero}):
\begin{eqnarray}
{d\phi_0\over d\hat t}&=&(g-1){Q\over2} +\nu(\hat t),\notag
\end{eqnarray}
where $\nu$ represents Gaussian noise. So $\phi_0(t)$ is a Wiener process with drift $(g-1)Q/2$,
\begin{eqnarray}
\phi_0(\hat t)&=&\phi_{0}(0)+(g-1){Q\over2}\cdot \hat t+\sqrt{\hat t}\cdot\epsilon,\notag
\end{eqnarray}
where the cumulative noise $\epsilon$ has variance 1. The area $A$ is thus a geometric Brownian motion.
Genus zero surfaces shrink exponentially in background time $\hat t$, until they reach a minimum area  $A_\text{min}\propto l^2$, where $l$ is a short-distance cutoff.
Surfaces of any fixed genus $g>1$ grow exponentially, until they eventually reach a maximum area $A_\text{max}\sim (g-1)/\mu$.\\

\begin{figure}[t]\centering
\includegraphics[height=5cm]{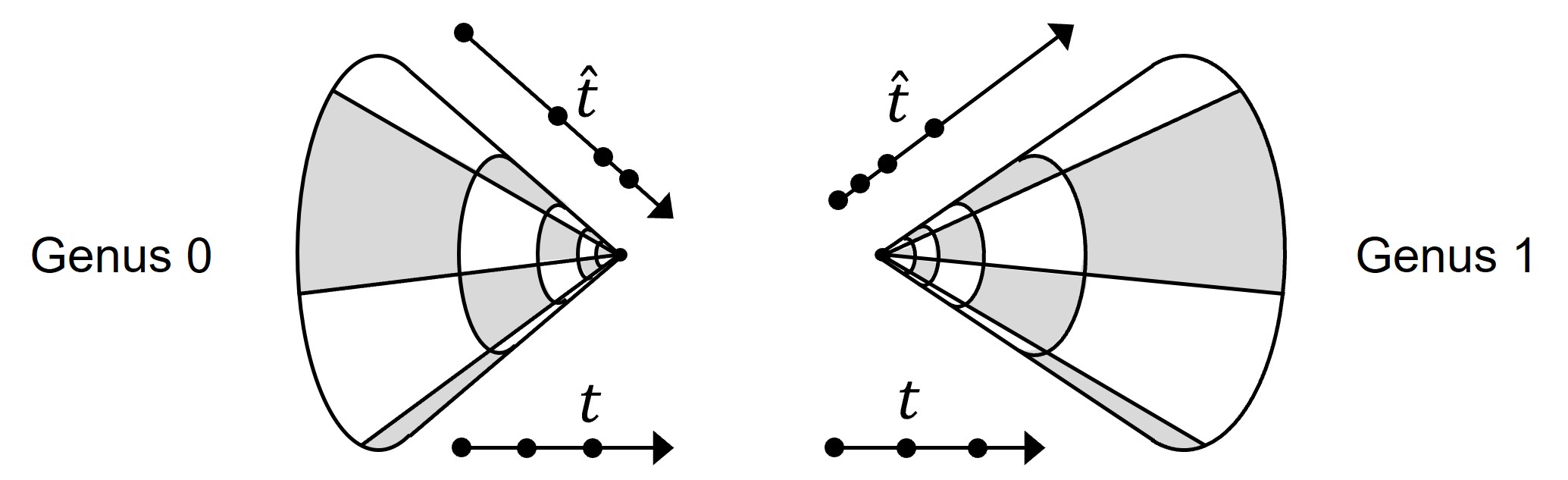}\hspace{1cm}
\caption{Left: small genus zero surfaces shrink linearly in physical time $t$ and exponentially in background time $\hat t$. Right: small surfaces of genus $g\ge1$ grow analogously.}
\end{figure}

How does the area evolve in {\it physical} time $t$?
From (\ref{time}), if $\hat t$ is fixed, $t$ is a random variable, and vice versa. 
Using the following identity implied by Ito's lemma, 
\begin{eqnarray}
\langle e^{\gamma\phi_0}\rangle&\sim&\exp\{{\gamma\over2}\big[\gamma+Q(g-1)\big]\cdot \hat t\}\ \ \ \text{for all}\ \ \gamma\in R,\notag
\end{eqnarray}
and setting $\gamma=\alpha$ yields the expectation value of physical time for genus $g=0$, using (\ref{cosmo}):
\begin{eqnarray}
{d\over d\hat t}\langle t\rangle\ \sim\ \langle e^{\alpha\phi_0}\rangle\ =\ e^{\omega\hat t}\ \ \ 
\text{with}\ \ \omega={\alpha \over2} \big(\alpha-Q\big)=-1\ \ \ 
\Rightarrow\ \ \langle  t_0-t\rangle\ =\ e^{-\hat t}\notag
\end{eqnarray}
with free parameter $t_0$. We see that physical time $t$ is exponentially related to background time $\hat t$. 
Background time $\hat t$ runs from $-\infty$ to +$\infty$, while physical time  $t$ runs from $-\infty$ to a finite end time $t_0$. 
This also implies a linear evolution of the area $t$: 
\begin{equation}
\langle A\rangle\sim\langle e^{\alpha\phi_0}\rangle \sim  e^{-\hat t}\sim\langle t_0-t\rangle.\notag
\end{equation}
So the area shrinks linarly in $t$, and the surface disappears at finite physical time $t_0$ (fig. 2, left) or shrinks to the cutoff size, if a minimum area cutoff is introduced.\\

The same calculation for any fixed genus $g\ge1$ shows that physical time $t$ runs from a finite starting time $t_0$ to $+\infty$. 
Genus $g\ge1$ surfaces are born at $t_0$, and their area grows linearly in physical time at rate $\omega_g =g\cdot Q\alpha/2-1$ (fig. 2, right).

\subsection{A Cox-Ingersol-Ross Process}

We can in fact include the cosmological constant $\mu$ and read off not just the expectation value, but the whole stochastic process of the area $A(t)\sim e^{\alpha\phi_0(t)}$ 
by restricting action (\ref{phys}) to the zero mode and changing variables from $\phi$ to $A$:
\begin{eqnarray}
 S_A&=&{1\over2}\int {dt\over A} \Big\{{1\over\alpha}\partial_t A+{Q\over2}(1-g)+{\mu\alpha\over2} A\Big\}^2-{\mu\alpha^2\over4}\int dt.
\notag\end{eqnarray}
Interestingly, a comparison with (\ref{dynaction}) shows that the area $A(t)$ follows a Cox-Ingersol-Ross process \cite{cir} for genus $g>0$: 
\begin{eqnarray}
{d\over dt}A&=&a(b-A)+\alpha\sqrt{A}\cdot\epsilon\ \ \ \text{with}\ \ \ a={\mu\alpha^2\over2},\ b={Q\over\mu\alpha}(g-1),
\label{cir}\end{eqnarray}
where $\epsilon$ represents the noise. This process is often
used in finance to model interest rates. It is also used in the Heston volatility model as a stochastic process for the variance \cite{heston}. 
In section 5, we will indeed show that it models the variance of time variations of the order parameter. 
The area mean-reverts to the equilibrium value $b$ at rate $a$. The factor $\sqrt A$ in front of the noise prevents the area $A$ from becoming negative for genus $g>1$.
The drift is $ab+\alpha^2/2$, where the noise term creates the additional drift $\alpha^2/2$ we encountered at the end of the previous subsection.
For genus zero, the drift is $-1$. As the area $A(t)$ approaches $0$, physical time stops growing (as $dt=A\cdot d\hat t$), so the paths $A(t)$ end at the boundary $A=0$. \\

The time-dependent probability distribution of the Cox-Ingersol-Ross process is known to be a non-central chi-squared distribution with $4ab/\alpha^2$ degrees of freedom. 
In the static limit for $g\ge1$, the probability density of the area $A$ approaches
the equilibrium distribution
$$\rho(A)\sim e^{-\mu A}\cdot A^{\nu-1}\ \ \ \text{with}\ \ \ \nu={2ab\over\alpha^2}={Q\over\alpha}(g-1)=(2+{1\over m})(g-1).$$
As a cross-check, this reproduces the fixed-area partition function (\ref{fixedA}). 

\subsection{Operator Insertions}

We can generalize this discussion to surfaces with operator insertions.
In the static limit, they have been discussed in \cite{seiberg,stau}. On a surface of genus $g$, consider the correlation function
$$\langle\prod_i e^{\alpha_i(\vec\sigma_i)}\rangle=Z^{-1}\int D\phi\ \exp\Big\{-{1\over8\pi}\int_\Sigma d^2\vec\sigma\sqrt{\hat g}\ \big( \partial\phi^2+Q\hat R\phi+\mu e^{\alpha\phi}\big)+\sum_i\alpha_i\phi(\vec\sigma_i)\Big\}$$
where $\hat R$ is the background Ricci scalar. The operator insertions are equivalent to curvature insertions 
both in the background metric and the physical metric: 
$$\sqrt{\hat g}\hat R(\vec\sigma)\ =\  -{8\pi\alpha_i\over Q}\cdot \delta(\vec\sigma-\vec\sigma_i)\ =\  \sqrt{g} R(\vec\sigma)+...$$
where the dots stand for terms involving spatial derivatives of $\phi$. 
The curvature singularities are pointlike in the background metric. 
Whether they are also pointlike (``microscopic'') in the physical metric, or cut holes into the surface (``macroscopic'') depends on $\alpha_i$ \cite{seiberg,stau}. \\

To compute these correlation functions, we can expand around classical solutions. Those are constant negative curvature solutions with these curvature singularities (fig. 3, left). 
They exist, if $\sum_i\alpha_i> Q(1-g)$, or if there is at least one boundary.
Otherwise, we must fix the area $A$ using a Lagrange multiplier. 
The classical solutions that we can expand around then have zero or constant postive curvature away from the insertions \cite{stau} (fig. 3, center).\\

\begin{figure}[t]\centering
\includegraphics[height=4.5cm]{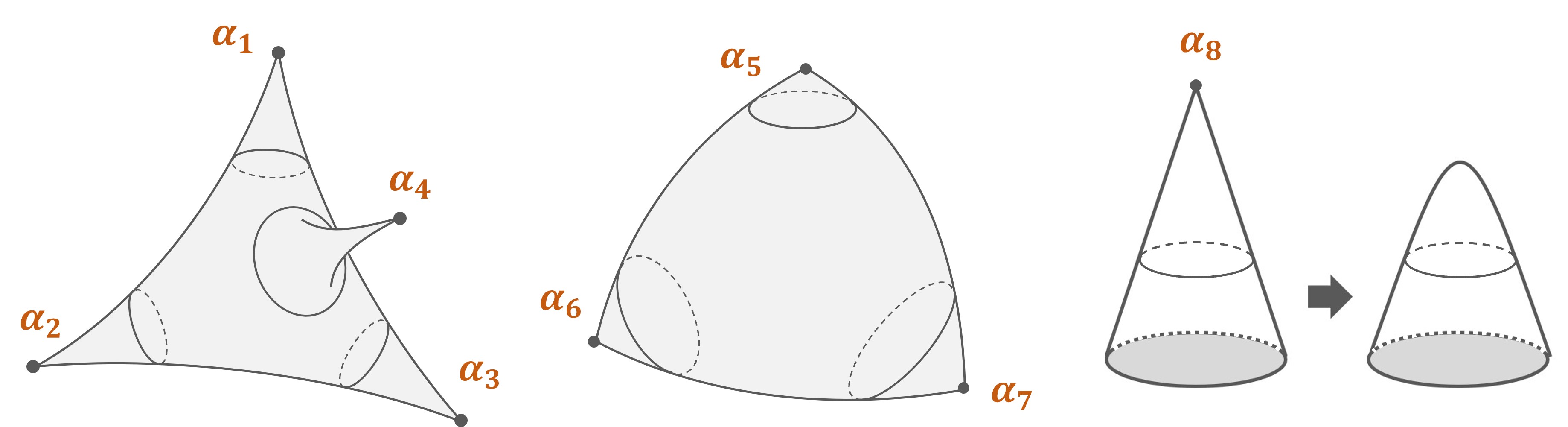}
\caption{Left and center: negative and positive curvature surfaces with operator insertions.
Right: dissipation of curvature in time after an operator insertion is removed.}
\end{figure}

What is the dynamics of these constant curvature surfaces with operator insertions? 
The Cox Ingersol Ross process of the previous subsection again applies, if we now choose a 
constant curvature background metric $\hat g$ with these pointlike curvature singularities. 
Then the $\phi$ zero mode decouples, and the stochastic process for the area is
$${d\over dt}A={\alpha \over2}\big[Q(g-1)+\sum_i\alpha_i\big]-{\mu\alpha^2\over2}A+\alpha\sqrt{A}\cdot\epsilon.$$
Thus, the area of these constant curvature surfaces shrinks to zero for positive curvature, 
corresponding to the case $Q(1-g)>\sum\alpha_i$, in which there is no static classical soution. 
For negative curvature, the area grows and asymptotically approaches
a maximum set by the cosmological constant (the static classical solution). \\

The above discussion assumes that the operator insertions remain on the surface at all times.
If instead we want to compute correlation functions of operators $e^{\alpha_i\phi(\vec\sigma_i,t_i)}$ at different points in time $t_i$,
then the curvature singularities are inserted only at times $t_i$ and dissipate thereafter. A numerical analysis indicates that both the area element $e^{\alpha\phi(\vec \sigma)}$ 
and the curvature $R(\vec \sigma)$ at the location $\vec \sigma$ of the cusp then decrease in physical time $t$ as $1/t$ (fig. 3, right).

\subsection{Correlation Functions}

Let us now discuss the $\phi$-zero mode contribution to time-dependent correlation functions of gravitational dressings.
To this end, we first introduce time boundaries: $\hat T_1<\hat t<\hat T_2$. 
Neglecting the cosmological constant (valid for small area $A$), the zero mode action (\ref{zero})  is
$$S_0={1\over2}\int_{\hat T_1}^{\hat T_2} d\hat t\big[\dot\phi_0+{Q\over2}(1-g)\big]^2
={1\over2}\int_{\hat T_1}^{\hat T_2} d\hat t\ \dot\phi_0^2+{Q\over2}(1-g)\big[\phi_0(\hat T_2)-\phi_0(\hat T_1)\big]$$
up to a constant. We see that the background charge amounts to inserting operators with opposite charges $\pm Q(1-g)/2$ at the time boundaries: up to a constant,
$$C^{(n)}(\hat t_1,...,\hat t_n)=\langle e^{\gamma_1\phi(\hat t_1)}\ ...\ e^{\gamma_n\phi(\hat t_n)}\rangle_Q
=\langle e^{+{Q\over2}(1-g)\phi({\hat T_1})}\ e^{\gamma_1\phi(\hat t_1)}\ ...\ e^{\gamma_n\phi(\hat t_n)}\  e^{-{Q\over2}(1-g)\phi({\hat T_2})}\rangle_{Q=0}.$$
The quantum mechanical propagator for the free field $\phi$ is
\begin{equation}
\Delta(\hat t_1,\hat t_2)=-{1\over2}\vert \hat t_1-\hat t_2\vert,\label{Coulomb}
\end{equation}
which yields the following result 
\begin{equation}
C^{(n)}(\hat t_1,...,\hat t_n)=\prod_{i<j}e^{-\gamma_i\gamma_j\vert \hat t_i-\hat t_j\vert}\cdot \prod_{k=1}^n e^{{\gamma_k Q}(1-g) ({\bar T}-\hat t_k)}
\ \ \ \text{with}\ \ \ \bar T={\hat T_1+\hat T_2\over2}.\notag
\end{equation}
Thus, in minisuperspace approximation and for small area, 
correlation functions of gravitational dressing operators are the free energy of a 1-dimensional Coulomb gas of particles with charges $\gamma_i$
in the presence of boundary charges $\pm Q/2$. 

\section{Dynamics of the Genus}

So far, we have kept the genus $g$ of the random surfaces fixed.
However, the genus in fact also evolves dynamically. We now study an extended minisuperspace with two dynamical variables $\phi_0$ and $g$.
In the static limit, their effective action from (\ref{gravity},\ref{anomaly}) is the potential
$$V(\phi_0,g)=(\ln \kappa^2+Q\phi_0)(1-g)+\mu e^{\alpha\phi_0}  + l^2 e^{-\alpha\phi_0} +\omega_g,$$
where $\omega_g$ comes from integrating over the moduli space of genus-$g$ surfaces, 
and we have added a small-area cutoff $l^2$, whose precise form should not matter; 
instead of a hard cutoff $A\sim e^{\alpha\phi_0}\ge l^2$, we suppress small areas by $e^{-V}\sim e^{-l^2/A}$, similarly as in (\ref{cutoff}).
Model A (\ref{modelA}) yields differential equations for changes in the expectation values of $\phi_0$ and the genus $g$: 
\begin{eqnarray}
\langle{\delta \phi_0\over\delta t}\rangle&=&-{1\over2}{\delta V\over\delta\phi_0}\ =\ {1\over2}Q(g-1)-{1\over2}\mu\alpha\ e^{\alpha\phi_0}+{1\over2}l^2\alpha\ e^{-\alpha\phi_0}\label{stophi}\\ 
\langle{\delta g\over\delta t}\rangle&=&-{1\over2}{\delta V\over\delta g}\ =\ {Q\over2}(\phi_0-\phi_c)\ \ \ \text{with}\ \ \ \phi_c={1\over Q}(2\omega'_g-\ln \kappa^2),
\label{stogenus}\end{eqnarray}
with constraint $g\ge0$.
For small area ($\phi_0\rightarrow-\infty$), the genus is driven to zero. For large area, it keeps growing, as shown in the flow diagram (fig. 4, left).
For large $\phi_0$, the flow of $\phi_0$ is halted by the cosmological constant $\mu$ in (\ref{stophi}), and for small $\phi_0$ by the cutoff $l^2$.
The figure shows the flow in the regime of intermediate $\phi_0$ and small $g$, neglecting $\omega_g$. \\

\begin{figure}[t]\centering
\includegraphics[height=5cm]{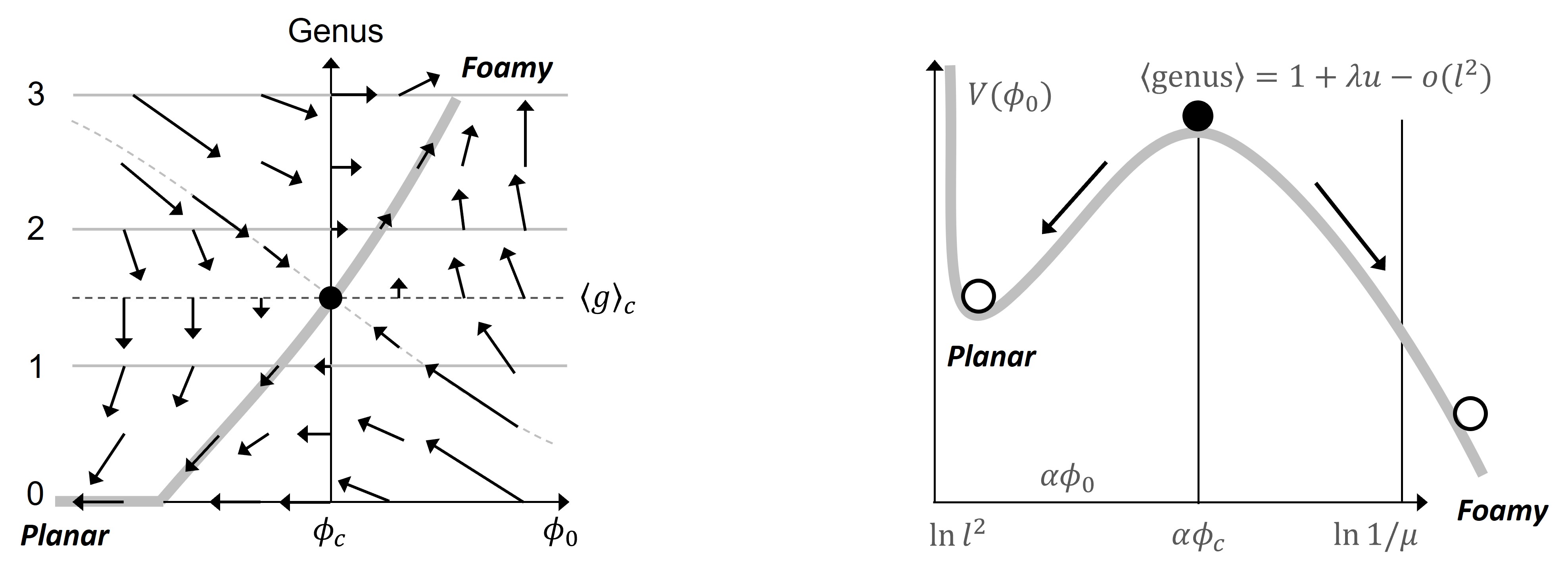}\vspace{4mm}
\caption{Left: dissipation of $\phi_0$ (the logarithm of the area) and of the genus $g$ of the random surface in time. 
Right: effective potential along the gray fixed line in the flow diagram.}
\end{figure}

Thus, depending on the initial value of $\phi_{0}$, there are two possible regimes. 
For small initial $\phi_{0}$, the system dissipates in time to small planar surfaces, rolling down the effective potential 
(shown in fig. 4, right, along the gray fixed line) to the left towards the small-area cutoff.
For large initial $\phi_{0}$, the system dissipates to large non-planar surfaces with a growing number of handles.
The two regimes are separated by an unstable fixed point with 
\begin{equation}
\langle g\rangle_c=1+\lambda u-{l^2\mu\over\lambda}\cdot{1\over u},\ \ \ 
\text{where}\ \ \ u\equiv \mu\cdot \kappa^{-2\alpha/Q},\ \ \ \lambda\equiv e^{2{\alpha\over Q}\omega'_g}.\label{uuu}
\end{equation}
Note that expectation value of the genus is approximately 1 for small $\mu$ and $l^2$. 

Does the effective potential have a second minimum in the non-planar regime, or do the area and the genus keep growing?
To answer this, one could try to compute the ground state energy and the expectation value $\langle g\rangle$ of the genus as a perturbation expansion in $g$.
If they converged, this would indicate an equilibrium distribution, i.e., a second minimum with finite average genus.
Unfortunately, these expansions diverge and are not Borel summable:
for a  given genus $g$, $V(\phi_0,g)$ has its minimum at area $A=Q(g-1)/(\alpha\mu)$, at which $\exp(-V)$ grows factorially with $g$.
This non-Borel-summability is a signal of instantons that give
a nonperturbative contribution to the free energy, invisible in the expansion in $\kappa$ \cite{ZJ}.\\

The nonperturbative free energy can actually be derived using the matrix models \cite{mxm,mxm2,mxm3,migdal}.
From (\ref{topo}), genus-$g$ surfaces are weighted by a power $(\kappa^2 A^{Q/\alpha})^{g-1}$ in $Z$, or, after integrating over the area,
by $(\kappa^2 \mu^{-Q/\alpha})^{g-1}$.
In terms of the modified inverse topological coupling constant $u$ defined in (\ref{uuu}),
the specific heat $ f(u)\equiv -Z''(u)$ of the $m$-th minimal model an a random surface satisfies generalizations of the Painlev\'e equation:

\noindent
\begin{eqnarray}
m=2:\ \ \ u&=&f^2-{1\over3}f''\ \Rightarrow\ f= u^{1/2}(a_0 +a_1u^{-5/2}+a_2u^{-5}+...)\label{painleve}\\
m=3:\ \ \ u&=&f^3-ff''-{1\over2}(f')^2+{2\over27}f'''',\ \ \ ...
\notag\end{eqnarray}
$m=2$ corresponds to random surfaces without matter, $m=3$ to the Ising model on a random surface, and so on. 
Choosing the boundary condition $f(u)\rightarrow u^{1/m}$ as $u\rightarrow\infty$ ($\kappa\rightarrow0$) and 
expanding in $1/u$ replicates the genus expansion, as shown in the case $m=2$. This allows us to derive 
the $\omega'_g$ in (\ref{stogenus}) from the $a_g$. 
As expected from the non-Borel-summability, (\ref{painleve}) also implies nonperturbative contributions to the free energy of the form 
\begin{equation}
c\cdot u^{\nu}\exp\{-B\cdot u^{{Q\over2\alpha}}\}\ =\ c\cdot u^{\nu} \exp\{-{B\over\kappa}\cdot\mu^{1+{1\over2m}}\}\label{tunnel}
\end{equation}
where $B$ and $\nu$ are numbers ($B={12\sqrt 3\over5}, {9\sqrt 2\over7},...$ and $\nu=-{1\over8},{1\over4},...$ for $m=2,3,...$), 
and $c$ is a new parameter that is invisible in the perturbation expansion in $\kappa$.
It has been argued in \cite{david,david2} that the unique physically acceptable solution for $f(u)$ has %no poles on the real axis but 
an imaginary coefficient $c$,
which signals that (\ref{tunnel}) is a tunneling amplitude. Indeed, in the matrix model (\ref{tunnel}) can be attributed to the tunneling of eigenvalues \cite{david2},
resulting in a phase where handles (long-range links) condense.
In terms of critical dynamics, it is natural to interpret this  
as tunneling from the planar to the foamy regime through the potential barrier in fig. 4.\\

The tunneling rate (\ref{tunnel}) beomes negligible if $u$ is large enough, 
i.e., if $\kappa$ is small enough or the cosmological constant is large enough. 
Then the planar regime is quasi-stable and we expect it to describe the ensemble of random surfaces well at large scales. 
However, in the case of small $u$ we 
conclude that the random surfaces are unstable in the sense that handles (and holes, if we allow for surface boundaries) 
become dense at the lattice scale. 
As all nodes are highly connected in such a "foamy" regime, we expect mean field theory to be exact.\\
 
To summarize, we expect two possible equilibrium states: a planar phase that gives rise to nontrivial critical phenomena and is described by the conformal field theory of section 2,
and a foamy phase that is described by mean field theory, and in which we do not expect nontrivial critical exponents. 
In the following, we will mostly focus on the planar phase.\\ 

As an illustration, fig. 5 (left) shows a typical planar random surface \cite{bettinelli}. It looks not unrealistic for a 
coarse-grained snapshot of a social network.
For comparion, fig. 5 (right) also shows an example of branched polymers, which might be related to matter with central charge $c>1$ 
or to certain non-unitary models with negative $c$ on a random surface.

\begin{figure}[t]\centering
\includegraphics[height=5cm]{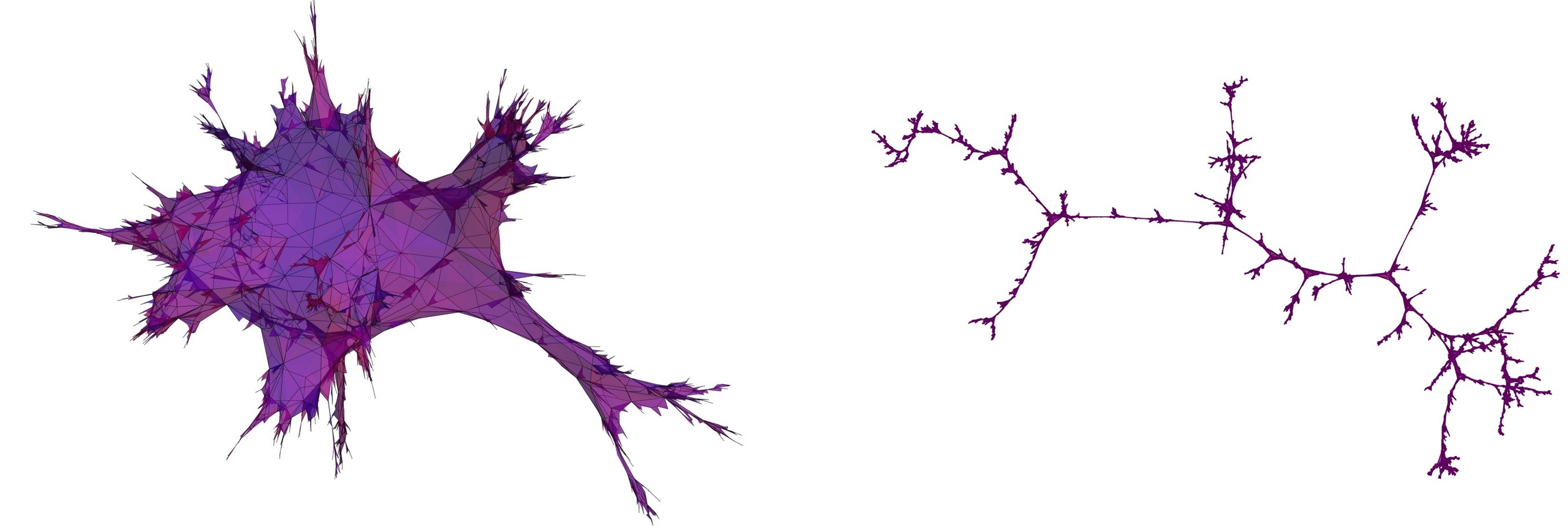}
\caption{Left: snapshot of a planar random surface; Right: snapshot of branched polymers. Source of images: home page of J. Bettinelli \cite{bettinelli}.} 
\end{figure}

\section{Dynamics of the Order Parameter}

In this section, we discuss variations of the order parameter over a given time interval $T$, which we call its ``returns''.
In the case of the Ising model, these returns are changes of the overall magnetization over this time interval.
They are important, because their distribution can be compared with empirically observed dynamic properties of real-world networks.

\subsection{Returns of the Order Parameter}

We begin with the minimal models without gravity. The operator $\pi(\hat t)$ at background time $\hat t$ represents the order parameter
$\Phi(\vec \sigma,\hat t)$ of the matter theory, such as the magnetization in the Ising model, integrated over the (static) surface $\Sigma$ of area $\hat A$:
\begin{equation}
\pi(\hat t)=\int_\Sigma d^2\sigma\ \Phi(\sigma,\hat t) .\notag
\end{equation}
In this section, we study the second moment $M_2(\hat T)$ of the distribution of ``returns'' of $\pi$, i.e., of its time variations over a given time horizon $\hat T$.
It is defined as
\begin{eqnarray}
M_2(\hat T)&=&\langle\big[\pi(\hat t+\hat T)-\pi(\hat t)\big]^2\rangle\ =\ \int_0^{\hat T} d\hat s \int_0^{\hat T} d\hat t\ \langle\dot\pi(\hat s)\dot\pi(\hat t)\ \rangle.\label{fixed}
\end{eqnarray}
The second moment is the variance of returns. On a flat surface $\Sigma$ of area $\hat A$, it is \cite{schmi2}
\begin{eqnarray}
M_2&\sim&\hat A\cdot \hat T^{{2\over z}(1-\Delta)}\ \ \ \text{for}\ \ \ \hat T\ll \hat A^z\label{varianz}
\end{eqnarray}
where $\Delta=\Delta_{22}$ is the dimension of $\Phi$ ($\Delta=1/8$ in the case of the Ising model).
The first factor $\hat A$ reflects translation invariance on the surface $\Sigma$. 
(\ref{varianz}) follows from the renormalization group by requiring the correct behavior under scale transformations 
$$\sigma\rightarrow\lambda\cdot\sigma,\ \hat A\rightarrow\lambda^2\cdot \hat A,\  \hat T\rightarrow\lambda^z\cdot \hat T,\ \pi\rightarrow\lambda^{2-\Delta}\pi,$$ 
as well as consistency with the limit case $\Delta=0, z=2$, which corresponds to an ordinary random walk with linearly growing variance $M_2\sim \hat T$.

\subsection{Time Evolution of the Variance}

Let us now couple the matter to gravity, where the area $A(\hat t)\sim\hat A e^{\alpha\phi(\hat t)}$ is dynamical.
Translation invariance on the surface suggests the following generalization of (\ref{varianz}):
\begin{eqnarray}
w_{\hat T}(\hat t)\ \equiv\ M_2(\hat T,\hat t)&=&A(\hat t)^\rho\cdot g(\hat T)\ \ \ \text{with}\ \ \ \rho=1,\label{ansatz4}
\end{eqnarray}
where $A(t)$ is the {\it physical} area that contains the time dependence. We will assume this ansatz here.
The dependence $g(\hat T)$ on the time horizon $\hat T$, higher moments, and corrections to $\rho$ 
that arise beyond the minisuperspace approximation will be discusssed in \cite{companion}. \\

\begin{figure}[t]\centering
	\includegraphics[height=4.5cm]{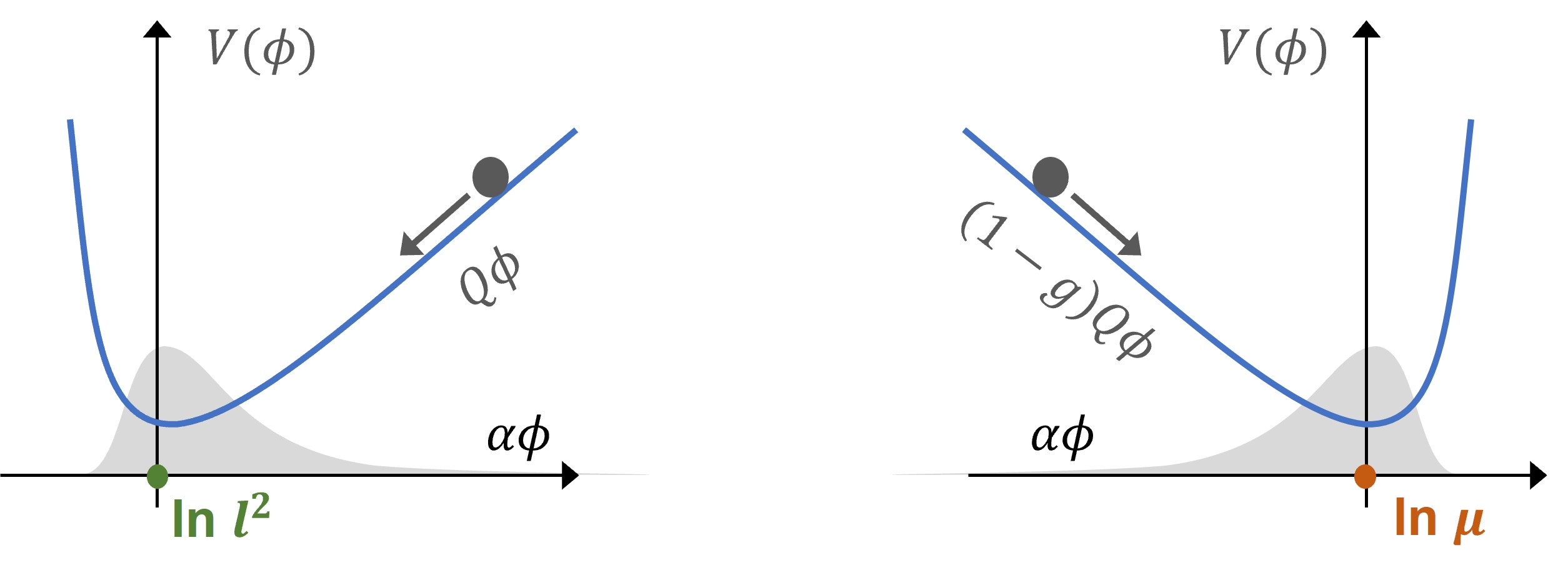}
	\caption{Left: for planar surfaces, the distribution of $\phi$ has a sharp lower bound. Right: for higher genus surfaces, the distribution of $\phi$ has a sharp upper bound.}
\end{figure}

Let us first consider the time evolution of the variance $w(\hat t)$ for fixed horizon $\hat T$. 
It follows from the evolution (\ref{stophi}) of the zero mode $\phi_0$ in its potential, assuming a fixed genus $g$:
\begin{equation}
w(\hat t)\sim e^{\rho\alpha\phi_0(\hat t)}\ \ \ \text{with}\ \ \ \dot\phi_0={Q\over2}(g-1) - {1\over2}\mu\alpha e^{\alpha\phi_0} + {1\over2}l^2\alpha e^{-\alpha\phi_0}+\nu(\hat t)\label{var1}
\end{equation}
with random noise $\nu$. 
In the planar regime of section 4, this potential for $\phi_0$ rises steeply to the left but slowly to the right (fig. 6, left). 
Since $\rho>0$, this leads to occasional spikes in the variance $w(\hat t)$, when the area of the random lattice becomes large. 
They decay to a volatility floor corresponding to $A\sim l^2$. For higher genus,
the picture is inverted: there is a volatility ceiling of order $\mu^{-1/2}$ with downside
spikes of the volatility (fig. 6, right).\\

(\ref{var1}) describes the evolution in {\it background} time. What is the evolution in {\it physical} time? 
For genus $g\ge0$, the cutoff $l^2$ can be neglected, as the ensemble is dominated by large-area surfaces. 
Then the evolution of the area and thus of the variance in physical time follows the Cox Ingersol Ross process (\ref{cir}). 
On the other hand, for genus zero, the cosmological constant can be neglected, as the ensemble is dominated by small-area surfaces. 
We can then define the inverse variance $\omega^{-1}\sim A^{-1}\sim e^{-\alpha\phi_0}$.
Since this just switches the sign of $\phi_0$ in our ansatz (\ref{var1}), the {\it inverse} area evolves as a Cox Ingersol Ross process in the planar regime, and the inverse variance is a power thereof.

\subsection{Equilibrium Distribution of Returns}

Let us now average over time periods that are much longer than the average length of the volatility clusters. 
This yields an equilibrium distribution of returns
$$R_{\hat T}(\hat t)={\tilde\pi(\hat t+\hat T)-\tilde\pi(\hat t)}.$$
For a given interval size $\hat T$, let us assume that the probability distribution of these returns at a point in time $\hat t$ 
is Gaussian with variance $w_A(\hat t)\sim A(\hat t)$, as in ansatz (\ref{ansatz4}).
Non-Gaussian corrections, which arise when going beyond the minisuperspace approximation, will be discussed in \cite{companion}. 
Averaging over time then turns the return distribution into a mixture of normal distributions with different variances.
Using the partition function (\ref{fixedA}) as a weight function, this mixed distribution is
\begin{eqnarray}
\rho(R)&\sim&\int_{A>l^2} dA \ A^{-1-(2+{1\over m})(1-g)}\cdot e^{-{l^2\over A}-\mu A}\cdot A^{-1/2} \exp\{-{R^2\over 2A}\},\label{ghyp}
\end{eqnarray}
where we include the small-area suppression by the factor $e^{-l^2/A}$ 
(in case the cutoff is implemented by a boundary, there is an additional power of $A$ as in (\ref{cutoff}).) 
(\ref{ghyp}) is a generalized hyperbolic distribution, which is defined as the following mixture of normal distributions:
$$f_{\nu,l^2,\mu}(x)\sim\int_0^\infty dw\ w^{-{\nu\over2}-{3\over2}}\cdot \exp\{-{l^2\over w}-\mu w\}\cdot \exp\{-{x^2\over2w}\}.  $$
In general, the tails of this distribution decay exponentially (including the cases $g\ge1$). However, for genus $g=0$, 
$\mu\rightarrow0$, and after choosing $l^2=\nu/2$ by a rescaling, we obtain a Student's t-distribution with $\nu$ degrees of freedom, which has power-law tails: 
\begin{eqnarray} 
f(x)\rightarrow\vert x\vert^{-\nu-1}\ \text{for}\ \vert x\vert\rightarrow\infty, &&\nu=2{Q\over\alpha}=4+{2\over m},\notag
\end{eqnarray}
where the last formula evaluates $\nu$ for the $m$-th unitary minimal model using (\ref{ddk}). To summarize, with ansatz (\ref{ansatz4}) in minisuperspace approximation,
time variations of the order parameter have generalized hyperbolic distributions.
In the planar regime, where $\mu$ can be neglected, those are Student's t-distributions with 4 or more degrees of freedom.\\

For genus $g>1$, where we can take $l^2\rightarrow0$ instead of $\mu\rightarrow0$, the resulting distributions are variance gamma distributions.

\section{Summary and Outlook}

We have studied the critical dynamics \cite{hohenberg} of the minimal models on a random surface based on ``model A''.
%using results from both Liouville theory and the matrix models. 
As in the static limit, it is important to distinguish between {\it background} scales, measured in the background metric $\hat g$ and background time $\hat t$,
and {\it physical} scales, measured in the physical metric $\hat g e^{\alpha\phi}$ and physical time $t$. Let us summarize our main results.

\begin{itemize}\addtolength{\itemsep}{0 pt} 
\item[(i)]
The evolution of the area $A(t)$ in physical time follows the Cox-Ingersol-Ross process (\ref{cir}).
For genus $g>0$, the area grows linearly at rate $\propto Q(g-1)$ while it is small, 
then converges to an equilibrium value proportional to $1/\mu$. The area of planar surfaces shrinks linearly in physical time.
They disappear, unless a minimum area $A_{min}$ is introduced, e.g., by adding a boundary of fixed length.
For surfaces with operator insertions with dressing $e^{\alpha_i\phi}$, the growth rate is replaced by $Q(g-1)+\sum_i\alpha_i$.  
\item[(ii)]
The time evolution of the genus of the surface is described by the differential equations (\ref{stophi},\ref{stogenus}). 
The genus tends to increase for large surfaces, leading to a "foamy phase", and to decrease for small surfaces, leading to a "planar phase". 
In the foamy phase, handles (corresponding to long-range links of the network) condense
and we expect mean field theory to be exact. In the planar phase, long-range links are irrelevant at large scales, and 
we expect observable nontrivial critical phenomena. 
\item[(iv)]
The time evolution of the variance $M_2$ displays upside spikes in the planar phase and downside spikes for higher-genus surfaces. 
In minisuperspace approximation, the equilibrium distributions of changes of the order parameter are generalized hyperbolic distributions.
They have power-law tails with tail indices of 4 or higher.
\end{itemize}

Many of the features we have found resemble empirical observations in financial markets.
The spikes of the volatility of returns of the order parameter resemble those of the VIX market volatility index.
The Cox-Ingersol-Ross process that describes the time evolution of the area and the volatility is familiar from the Heston volatility model.
And the Student's t-distributions of returns with 4 or more degrees of freedom have proven useful to model financial market returns. 
In the follow-up paper \cite{companion}, we study the higher moments of the returns of the order parameter and show that they exhibit multifractal scaling
\cite{mandel,bacry},
which has also been observed in financial markets (see \cite{bouch2, tizi} for reviews).\\

These observations seem to support the proposal \cite{schmi2} of modeling efficient financial markets as a lattice gas that is driven to its critical point by arbitrageurs,
with ``price-minus-value'' in the role of the order parameter.
More generally, they point to a potential new application of conformal field theories on a random surface, namely as large-scale limits of certain social networks
that have a built-in mechanism of self-organized criticality \cite{bak}.\\

\section*{Acknowledgements} 

I would like to thank Henriette (formerly Wolfgang) Breymann, Uwe Täuber, Matthias Staudacher, Jean-Philippe Bouchaud, Ashkan Nikeghbali,  
Sara Safari, Thomas Lehérici, and Maximilian S. Janisch for interesting discussions.
This research is supported by the Swiss National Science Foundation under Practice-to-Science grant no. PT00P2\_206333.

\end{document}